\documentclass[onecollarge,final]{svjour2}

\usepackage{graphicx}
\usepackage{rotating}
\usepackage{amssymb}
\usepackage{mathptmx}
\usepackage[numbers]{natbib}
 \makeatletter

\begin{document}

\newcommand{\hdblarrow}{H\makebox[0.9ex][l]{$\downdownarrows$}-}
\title{Spin effects in Bose-Glass phases}

\author{S. Paganelli$^1$ \and M. \L{}\c{a}cki$^2$ \and V. Ahufinger$^3$\and J. Zakrzewski$^{2,4}$ \and A. Sanpera$^{5,1}$}

\institute{1: Grup de F\'{i}sica Te\`orica: Informaci\'{o} i Fen\`omens Qu\`antics, 
Universitat Aut\`onoma de Barcelona, 08193 Bellaterra, Spain\\
\email{paganelli@ifae.es}\\
2: Instytut Fizyki imienia Mariana Smoluchowskiego, Uniwersytet Jagiello\'nski, \\
ulica Reymonta 4, 30-059 Krak\'ow, Poland\\ 
3: Grup d'\`{O}ptica: Departament de F\'{i}sica Universitat Aut\`onoma de Barcelona, \\ 
08193 Bellaterra, Spain\\
4: Mark Kac Complex Systems Research Center, Jagiellonian University,Krak\'ow, Poland\\
5: ICREA-Instituci\'o Catalana de Recerca i Estudis Avan\c{c}ats,\\
Llu\'{i}s Companys 23, 08010 Barcelona, Spain
}

\date{\today}

\maketitle


\begin{abstract}
We study the mechanism of formation of Bose glass  (BG) phases in the spin-1 Bose Hubbard model when 
diagonal disorder is introduced. To this aim, we analyze first the phase diagram in the zero-hopping limit, there 
disorder induces superposition between  Mott insulator (MI) phases with different filling numbers. 
Then  BG appears
as a compressible
 but still insulating phase.
The phase diagram for finite hopping is also calculated with the Gutzwiller approximation.
The bosons' spin degree of freedom introduces another scattering channel in the two-body interaction modifying the stability of 
MI regions with respect to the action of disorder.
This leads to some peculiar phenomena such as the creation of BG of singlets, for very strong spin correlation, or the 
disappearance of BG phase in some particular cases where fluctuations are not able to mix different MI regions.

\end{abstract}

\section{Introduction}\label{sec:intro}
Ultracold atomic gases can be described by interacting atoms with an internal spin, 
corresponding to a low-energy hyperfine level $F$.
If the spin orientation is fixed by an external magnetic field, as it happens when the gas is confined in
 magnetic trap,  a 
scalar model is sufficient to describe the system.
Conversely, if the spin orientation is not externally constrained, as in the case of optical trapping,
the spinor character of the gas has to be taken into account.
For bosons trapped in a deep optical lattice potential  the system is well described by the spinor Bose-Hubbard (BH) 
model \cite{Imambekov03}.
Bosonic interactions, treated as two-body contact collisions, are sensitive to the spin degree of 
freedom and contribute to the orderings at zero temperature.

As in the scalar case \cite{Fisher89}, the competition between hopping and interactions
leads to a quantum phase transition between spinor superfluid (SF) condensate and a Mott insulator (MI) state
\cite{Imambekov03,Tsuchiya04,Rizzi05}. Spin correlation introduces magnetic ordering 
which contribute to the stability of one phase with respect to the other.
Moreover, the presence of spin scattering channels  influences also the stability of the 
MI phases in the presence of  different types of disorder.

Disorder plays an essential role in condensed matter physics and it has been shown to be
 an essential ingredient for studies of  conductivity, transport, high-Tc superconductivity, 
neural networks or quantum chaos to mention few examples 
(see the review\cite{Lewenstein07} and references therein). 
Disorder can be produced in ultracold atoms 
 in a {\it {controlled}} and {\it {reproducible}} way. Standard methods to achieve such a  controlled disorder 
are the use of speckle patterns \cite{Horak98,Boiron99} which can be added to the confining potential, or optical superlattices 
created by the simultaneous presence of optical lattices of incommensurate frequencies
\cite{Roth03,Damski03,Dinier01}. Other methods include 
using an admixture of different atomic species randomly trapped in sites distributed across the sample and acting as impurities 
\cite{Gavish05,Massignan06},
or the use  of inhomogeneous magnetic fields 
which modify randomly, close to a Feshbach resonance, the scattering length of atoms in the sample depending 
on their spatial position \cite{Gimperlein05,Chin10}.

Recently,  the phase diagram of the spin-1 BH model in two dimensions (2D) 
in the presence of  disorder has been studied with a Gutzwiller mean field 
approximation\cite{Lacki11}. 
As in the scalar case \cite{Fisher89}, a gapless Bose-glass (BG) insulator phase appears,   
characterised by  finite compressibility and exponentially decaying superfluid correlations in space, 
but, because of the spin interaction, the phase diagram changes considerably.

In this paper we focus on the spin-1 BH model in the presence of diagonal disorder and analyse the 
role of spin correlations in the formation of the BG phase.
We study in detail the zero-hopping limit (atomic case). We provide, in this simple case, the phase diagram 
corresponding to different types of disorder and show   how  spin correlations can prevent the formation of 
the BG between some MI regions. This fully analytical approach allows to easily visualise the 
mechanism of formation of the BG in terms of superpositions between MI phases with different 
filling factors. 
We also provide, in some cases, the complete phase diagram for finite tunnelling using a numerical
mean field Gutzwiller approximation. In particular, we show the case of large spin interactions where 
a BG of singlets emerges. We  analyse the case in which the disorder is directly introduced in 
the two-body scattering lengths, associated with total spin of scattering particles $s=0$ and $s=2$,
instead of in the Hamiltonian parameters directly.
This represents a more realistic scenario since small fluctuations in the scattering lengths
can be introduced by optical Feshbach resonances.
We observe the absence of BG phase for ferromagnetic spin interactions
and disorder in the $s=2$ scattering length, confirming that, in this limit, the spinor model is
equivalent to the scalar case.

  
The paper is organised as follows: in Sec. \ref{sec:model} we introduce the BH model. Then we proceed 
to compare the effect of diagonal disorder in the local potential or in the interaction.
In particular, we compare the cases in which  disorder is in the local parameters of the Hamiltonian ( Sec. \ref{sec:disdiag})
with the case in which disorder  is introduced directly in the two-body scattering lengths corresponding
to the collision channel  with total spin $s=0$ and $s=2$ (Sec. \ref{sec:scat}).
Finally, in Sec. \ref{sec:conc} we present our conclusions. 

\section{Model}\label{sec:model}
\
Low energy spin-1 bosons loaded in optical lattices, sufficiently deep so that only the lowest energy band is 
relevant, can be described by the spinor BH model.  The corresponding Hamiltonian is \cite{Imambekov03}:
\begin{equation} \label{eqn:BHham}
\hat{H}  =  -t\sum_{\left\langle i,j\right\rangle ,\sigma} \hat{a}_{i\sigma}^{\dagger}\hat{a}_{j\sigma}
+ \sum_{i} \left[ \frac{U_{0}}{2}\hat{n}_{i}(\hat{n}_{i}-1)+
\frac{U_{2}}{2}\left(\hat{\mathbf{S}}_{i}^{2}-2\hat n_{i}\right)-\mu \hat{n}_{i} \right],
\end{equation}
where $\left\langle i,j\right\rangle$ indicates that the sum is restricted to nearest neighbours in the lattice and 
$\hat{a}_{i\sigma}^{\dagger}$ ($\hat{a}_{i\sigma}$) denotes the creation (annihilation) operator of a boson in the lowest
Bloch band localised on site $i$  with spin component $\sigma=0,\pm 1$. 

The first term in (\ref{eqn:BHham}) represents the kinetic energy and describes spin independent hopping between 
nearest-neighbour sites with tunnelling amplitude $t$. The second and third term account for spin 
independent and spin dependent on site interactions, respectively. These energies at site $i$ are defined as 
$U_{0,2}=c_{0,2}\int d\vec{r} w^4(\vec{r}-\vec{r}_i)$ 
with $c_{0}=4 \pi\hbar^2(a_0+2a_2)/(3m)$  and $c_{2}=4 \pi \hbar^2 (a_2-a_0)/(3m)$, where $a_S$ with $S=0,2$ is the s-wave scattering 
length corresponding to the channel with total spin $S$ \cite{Ho98,Ohmi98} and $w(\vec{r}-\vec{r}_i)$ is the Wannier function of the 
lowest band at site $i$.  While the second term of (\ref{eqn:BHham}) is spin independent and 
equivalent to the interaction energy for
scalar bosons, the third term represents the energy associated with spin configurations within lattice sites with
\begin{equation}
\hat{\mathbf{S}}_i=\sum_{\sigma \sigma^{\prime}=0,\pm 1} \hat{a}^{\dagger}_{\sigma i} 
\vec{F}_{\sigma \sigma^{\prime}} \hat{a}_{\sigma^{\prime} i} ,
\label{si}
\end{equation}
being the spin operator at site $i$
 and $\vec{F}$ the traceless spin-1 matrices. 
$\hat{\mathbf{S}}$'s components obey standard angular momentum commutation relations $[\hat{S}_{l}, \hat{S}_{j}]=i\epsilon_{ljk} \hat{S}_{k}$.
The spin-interaction term favours a configuration with total magnetisation zero 
for $U_2>0$, denoted as polar and sometimes antiferromagnetic. The ferromagnetic 
configuration where spins add to a maximal possible value  corresponds to $U_2<0$ 
\cite{Ho98,Ohmi98,Law98}.
In the grand canonical approach the total number of particles is 
controlled by the last term of (\ref{eqn:BHham}) where $\mu$ is the chemical potential and  
\begin{equation}
\hat{n}_{i}=\sum_{\sigma=0,\pm1}\hat{n}_{i,\sigma},
\end{equation}                                           
is the total number of bosons on site  $i$. 
Hamiltonian (\ref{eqn:BHham}) can be straightforwardly derived from the microscopical description of bosonic atoms, with a 
hyperfine spin $F=1$, loaded in a deep optical lattice and considering the 
two-body short range (s-wave) collisions.  More details about the derivation can be found in 
\cite{Tsuchiya04,Ho98,Ohmi98,Jaksch98,Koashi00}. 
 
Notice also that, since the orbital part of the wave function in one lattice site is the product of
 Wannier functions for all the atoms, it is symmetric under permutation of any two atoms. Therefore, the spin part of the
 wavefunction should also be symmetric due to Bose statistics. This imposes  $s_i+n_i$ to be even \cite{Ying96}, 
being $s_i$ and $n_i$ the quantum numbers labelling the eigenvalues of $\hat{\mathbf{S}}_i$ and $\hat{n}_{i}$, respectively. 
In alkaline atoms usually, the scattering lengths are similar, $a_0\simeq a_2$, and the symmetry of the Hamiltonian becomes 
SU(3) instead of SU(2). That implies  $|U_0|\gg|U_2|$.

As in the scalar case, the spinor BH system exhibits a quantum phase transition between superfluid and 
insulating states \cite{Imambekov03,Tsuchiya04}.  In the insulating states, fluctuations in the atom number per site are suppressed 
and virtual tunnelling gives rise to effective spin exchange interactions that determine a rich phase diagram in which 
different insulating phases differ by their spin correlations. 
\subsection{Atomic limit ($t=0$)}
Hamiltonian (\ref{eqn:BHham}) becomes diagonal in the limit of vanishing hopping. 
In this limit it reduces to the sum of local terms  $\hat{H}\rightarrow\sum_i \hat{H}_{0,i}$ with
\begin{equation}
\hat{H}_0= \frac{U_{0}}{2}\hat{n}(\hat{n}-1)
+\frac{U_{2}}{2} \left(\hat{\mathbf{S}}^{2}-2\hat n\right)-\mu \hat{n},
\end{equation}
(site indexes are omitted if the system is homogeneous).
Eigenstates of $\hat{H}_0$ are denoted with $\left|n,s,m\right\rangle$, where the quantum numbers
refer to the three commuting local observables 
\begin{eqnarray*}
\hat{n}\left|n,s,m\right\rangle  & \mbox{=} & n\left|n,s,m\right\rangle, \\
\mathbf{\hat{S}}^{2}\left|n,s,m\right\rangle  & \mbox{=} & s(s+1)\left|n,s,m\right\rangle, \\
\hat{S}_{z}\left|n,s,m\right\rangle  & \mbox{=} & m\left|n,s,m\right\rangle, 
\end{eqnarray*}
with energies $H_0\left|n,s,m\right\rangle = E_{0}(n,s)\left|n,s,m\right\rangle $ given by
\begin{equation}\label{eqn:e0}
E_{0}(n,s)=-\mu n+\frac{U_{0}}{2}n(n-1)+\frac{U_{2}}{2} \left( s(s+1)-2n \right).
\end{equation}
From Eq.(\ref{eqn:e0}), one can easily deduce the structure of the ground state of the insulator phases in the limit $t=0$. 
In the atomic limit without disorder, only MI phases exist. SF corresponds to the points separating
MI intervals with different fillings (as can be inferred from $t\rightarrow 0$ limit).
This means that, at fixed $\mu$, the ground state has an integer filling $n$. The boundaries corresponding to the degeneracy
points between two fillings $n_1$ and $n_2$ satisfy the condition $E_0(n_1,s_1)=E_0(n_2,s_2)$.
Any MI region with filling $n$ and fixed $U_2$  is defined for $\mu_-(n)< \mu < \mu_+(n)$, where $\mu_\pm(n)$ are the 
boundaries of the region.

For antiferromagnetic interactions, $U_2>0$, the minimum energy $E^{min}_0$  is 
attained with minimum $s$, its specific 
value depending of the number of atoms per site. Thus, for the even filling factor, 
the minimum spin is zero and the state is 
described as $\left|0,0,n\right\rangle$ with $n$ even.
This state is known as spin singlet insulator \cite{Demler02}. 
If the atom number per site is odd, then the minimum spin per 
site is one and the state reads $\left|1,m,n\right\rangle$ in the absence of disorder.

Let us calculate explicitly $\mu_\pm$. For $U_2>0$ (antiferromagnetic case) we must distinguish between odd and even occupation lobes: 
 
\noindent
(1) $n$ odd: the boundaries $\mu_\pm(n)$ are obtained by imposing $E_0(n,1)=E_0(n \pm 1,0)$
 \begin{eqnarray}\label{eqn:bound1}
   \mu_- &=&(n-1)U_0, \nonumber \\
   \mu_+ &=& n U_0-2 U_2.
 \end{eqnarray}
Odd lobes exist only for $\mu_-<\mu_+$ that is, for $U_2/U_0< 0.5$. 

\noindent
(2) $n$ even: for $U_2/U_0< 0.5$, $\mu_\pm(n)$ are obtained by imposing $E_0(n,0)=E_0(n \pm 1,1)$
 \begin{eqnarray}\label{eqn:bound2}
   \mu_- &=&(n-1)U_0-2 U_2, \nonumber \\
   \mu_+ &=& n U_0,
 \end{eqnarray}
while, for $U_2/U_0> 0.5$ the condition $E_0(0,n)=E_0(0,n \mp 2)$ leads to
 \begin{eqnarray}\label{eqn:bound3}
   \mu_- &=& \left(n-\frac{1}{2}\right)U_0-U_2, \nonumber \\
   \mu_+ &=& \left(n+\frac{1}{2}\right)U_0-U_2,
 \end{eqnarray}

For $U_2<0$ (ferromagnetic case) the maximum value of the spin is given by $s=n$ and the condition for the boundaries 
$E_0(n,n)=E_0(n \mp 1,n)$ reads
 \begin{eqnarray}\label{eqn:bound4}
   \mu_- &=& \left(n-1\right)(U_0+U_2), \nonumber\\
   \mu_+ &=&  n (U_0+U_2).
 \end{eqnarray}

\begin{figure}
\begin{center}
\includegraphics[width=0.55\linewidth,keepaspectratio,angle=-90]{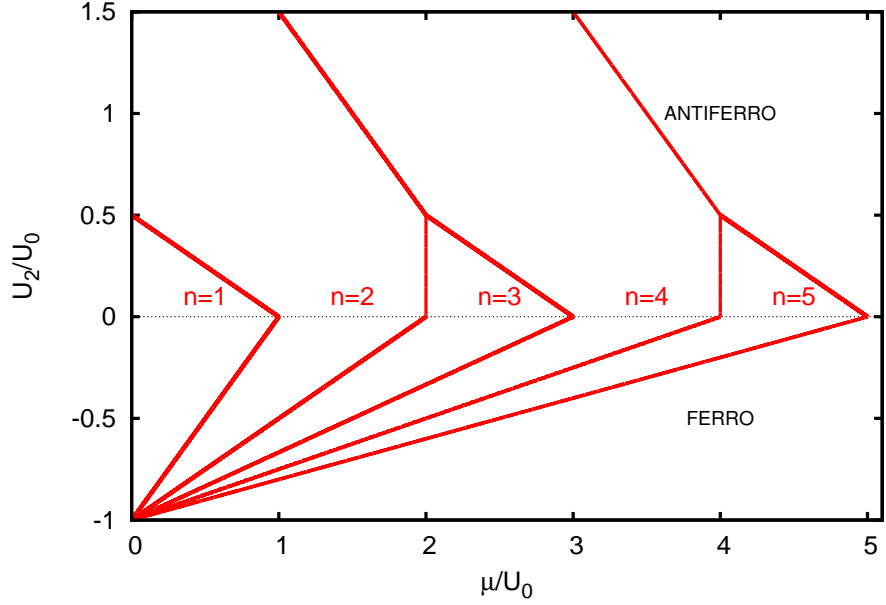}
\end{center}
\caption{Phase diagram of the spinor $F=1$ BH model in the limit $t=0$. Each region corresponds to a MI
phase with a different occupation number. For $U_2/U_0>0$ the system has zero 
magnetization (antiferro) and for  $U_2/U_0<0$ the system is in a ferromagnetic phase}
\label{fig:t0}
\end{figure}

The atomic-limit phase diagram is depicted in  Fig. \ref{fig:t0}, the MI intervals obtained fixing $U_2$
correspond to the  basis of the MI lobes in the $t/U_0$ - $\mu/U_0$ plane\cite{Lacki11}. 
In the antiferromagnetic region, one can see that the formation of singlet 
stabilises  the even MI lobes while the odd lobes shrink. 
For $U_2<0$, $U_2$, eq.  (\ref{eqn:BHham}) reduces to the scalar Hamiltonian with $U_0+U_2$ 
put in place of $U_0$. 
Notice that for $U_2<-U_0$ the spectrum (\ref{eqn:e0})  is not bounded from below and 
the model becomes instable.
 
\section{Diagonal disorder in the Hamiltonian's parameters}\label{sec:disdiag}

Starting from this scenario, we proceed to analyse the stability of the MI phase in the presence of disorder.
We start by analysing local fluctuations $\epsilon_i$  added either to the homogeneous chemical potential $\mu$, 
or on the interaction potential $U_0$ or $U_2$. 
We consider $\epsilon_i$ to be a random variable defined for every site $i$  with a
 given probability distribution $p(\epsilon)$.
Here we consider a bounded probability distribution that is $-\Delta < \epsilon_i < \Delta$. 
Disorder mixes together different MI regions so that degeneracy between different fillings appears.
So, between MI intervals, regions can appear where filling is not defined and correspondingly a  BG phase appears.
Being the probability distribution bounded, the BG phase appears only around the original boundaries and its extension depends on 
$\Delta$. On the other hand, deep inside the MI, small fluctuations 
of the disordered parameter are not able to mix different fillings and the 
state remains stable in the corresponding MI phase.

\begin{figure}
\begin{center}
\includegraphics[width=0.55\linewidth,keepaspectratio,angle=-90]{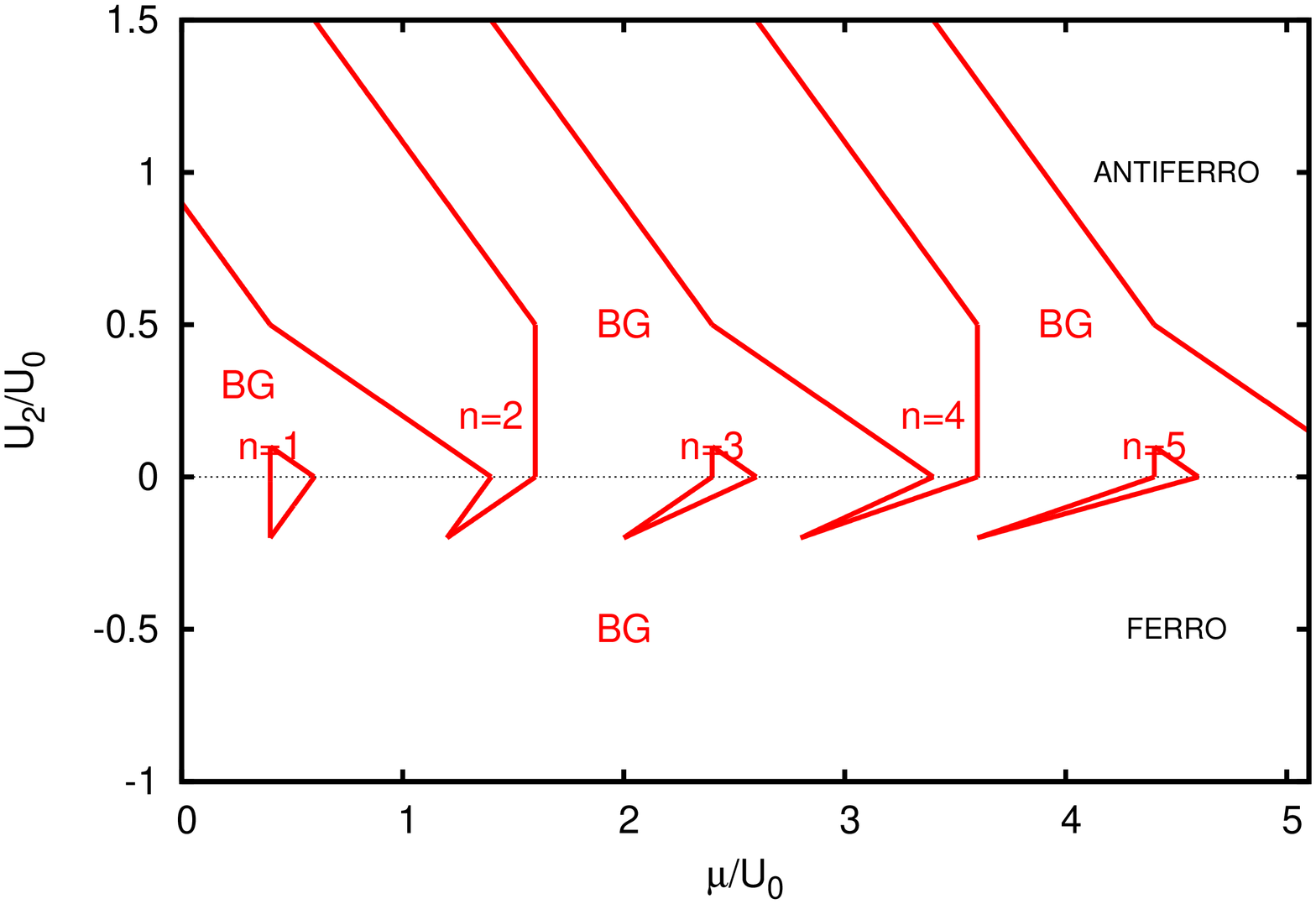}\\
\includegraphics[width=0.55\linewidth,keepaspectratio,angle=-90]{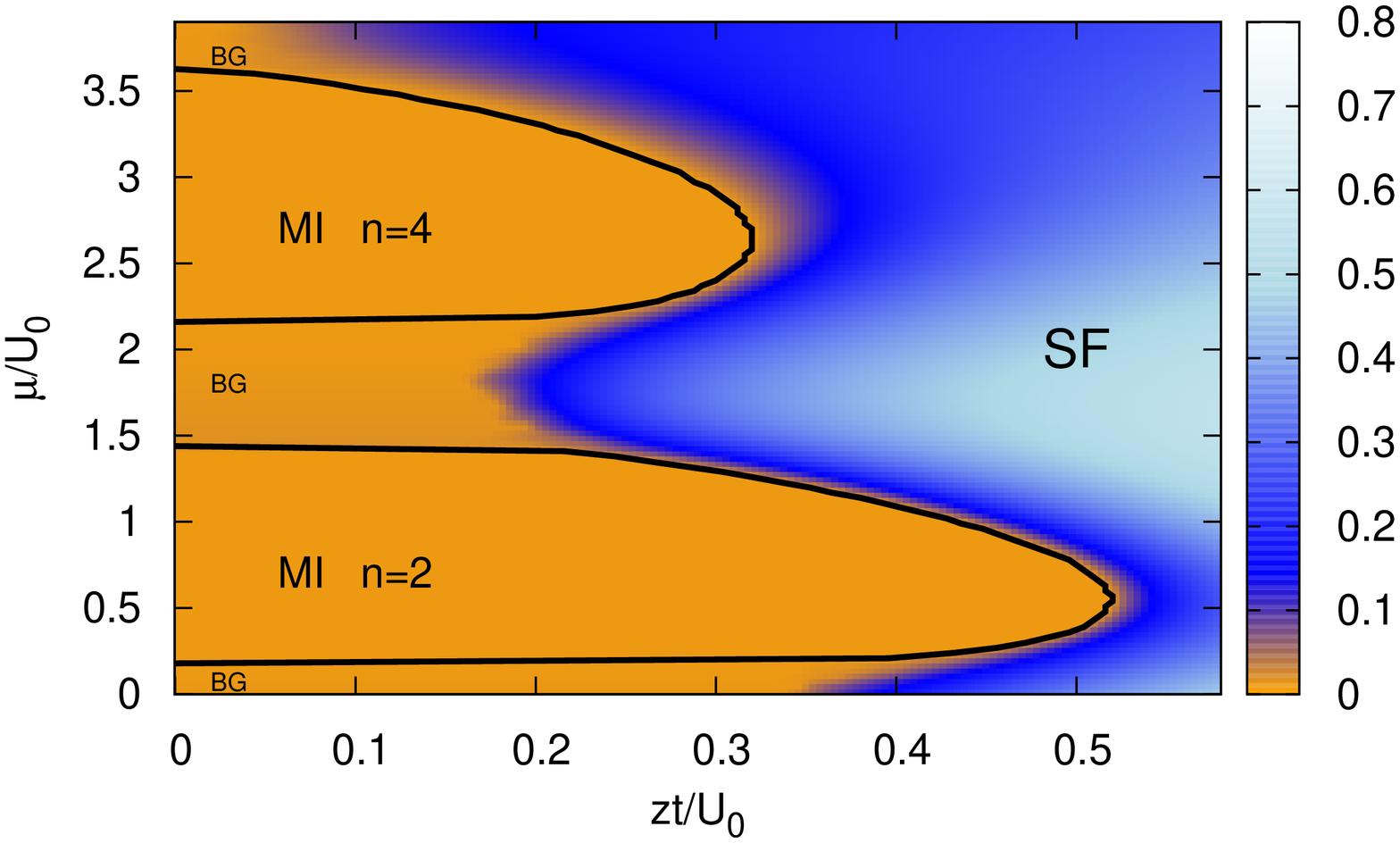}
\caption{(Color online) Top panel:  phase diagram in the limit $t=0$ and disorder in $\mu$ with  $\Delta/U_0=0.4$.
Bottom panel: condensate fraction with disorder in $\mu$ for $U_2/U_0=0.7$  and $\Delta/U_0=0.4$. 
The solid lines correspond to the boundaries of the MI lobes. The region outside the MI phase and with vanishing
condensate fraction corresponds to the BG. In this case BG is formed by singlets.}
\label{fig:t0_dismu}
\end{center}
\end{figure}

The first case we study is the disorder in the chemical potential. 
This type of disorder can be produced introducing some random inhomogeneities in the local potential 
$\mu_j= \mu + \epsilon_j$.
Fixing $U_2$, disorder in $\mu_j$ corresponds to horizontal fluctuations of maximum amplitude $\Delta$
in the atomic-limit phase diagram.
The new boundaries for the MI phases are given by replacing $\mu_\pm \rightarrow \mu_\pm \mp \Delta$ as can be observed in
the top panel of Fig. \ref{fig:t0_dismu}.
We notice that BG regions always appear between MI lobes with width (in $\mu$) of the order of $2 \Delta $.
Since odd lobes  shrunk by $U_2$, they are more unstable and disorder can make them disappear.

When $U_2/U_0>0.5$ disorder mixes together only 
even occupations so BG is formed by singlets \cite{Lacki11}. The finite-hopping phase diagram for this case is 
displayed in the bottom panel of Fig. \ref{fig:t0_dismu}, where 
we have used the condensate fraction \cite{Roth03} as an order parameter to separate SF from BG phase. 
Our calculations are done within the mean field Gutzwiller approach, which is able to take into account 
inhomogeneties caused by disorder\cite{Lacki11}. The MI phase is characterised
by vanishing fluctuations in the density (or zero compressibility) while the BG corresponds
to finite density fluctuations and zero condensate fraction. 
In this plot one can see the disappearance of the odd occupation MI lobes. Numerical calculation shows that
$\left\langle \mathbf{S}^{2} \right\rangle =0 $ meaning that also  the  BG phase is formed by singlets\cite{Lacki11}.  
Note that, even if we found it adding disorder in the chemical potential, singlet BG appears also 
for disorder in $U_0$ and $U_2$, the only necessary condition being $U_2/U_0>0$.

\begin{figure}
\begin{center}
\includegraphics[width=0.55\linewidth,keepaspectratio,angle=-90]{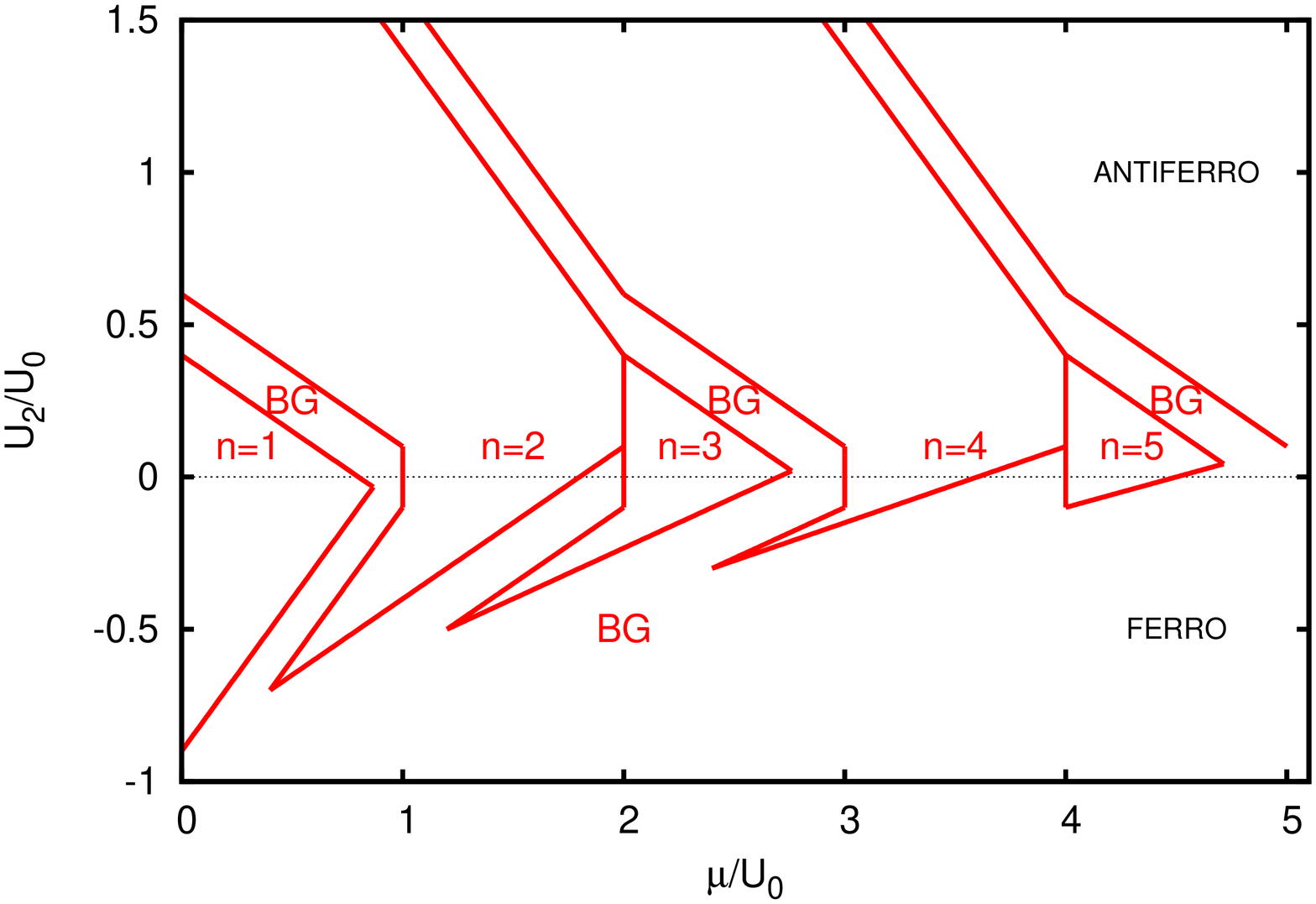}\\
\includegraphics[width=0.55\linewidth,keepaspectratio,angle=-90]{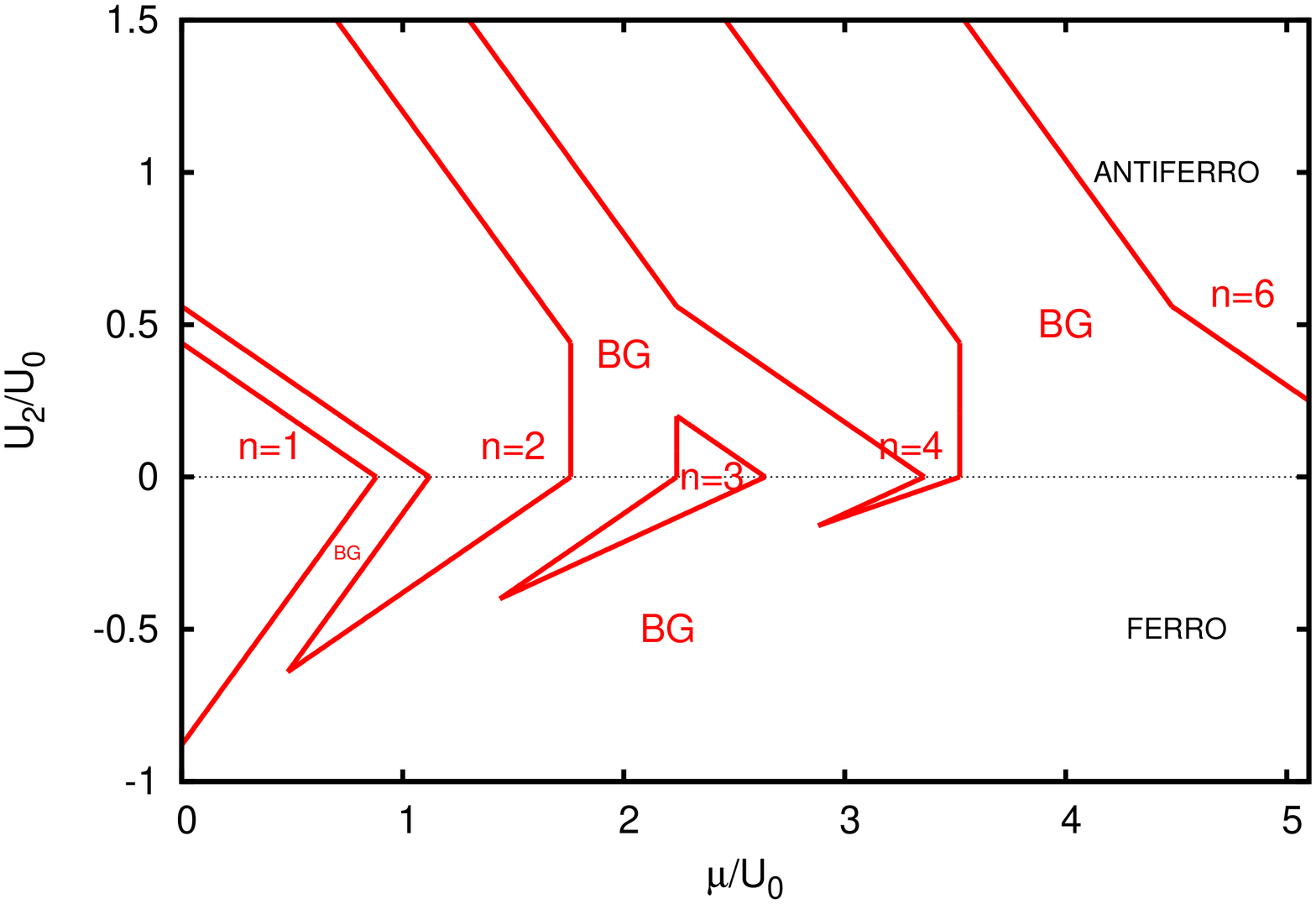}
\caption{Top panel: phase diagram in the limit $t=0$ and disorder in $U_2$ with $\Delta/U_0=0.1$.
Bottom panel:  phase diagram in the limit $t=0$ and disorder in $U_0$ with $\Delta/U_0=0.12$.}
\label{fig:t0_disU2U0}
\end{center}
\end{figure}

Disorder in $U_2$ corresponds to fluctuations along the vertical direction of the phase diagram
in Fig. \ref{fig:t0}.  The resulting phase diagram in the atomic limit is shown in the top panel of 
Fig. \ref{fig:t0_disU2U0}.
The new boundaries are obtained changing $\mu_\pm$ in Eqs. (\ref{eqn:bound1}-\ref{eqn:bound4})
replacing  $U_2 \rightarrow U_2 \mp \Delta$ for $U_2/U_0 <0$, and
$U_2\rightarrow U_2 \pm \Delta$    for $U_2/U_0 >0$ .
One consequence is that, for $U_{2}/U_0<0.5$  no BG appears between even and odd lobes with higher
 occupation (that is between lobes with occupation $n=2m$ and $n=2m+1$)  where the 
boundaries are vertical. 
This is true for $|U_2|>\Delta$ otherwise disorder can mix together ferromagnetic 
and antiferromagnetic regions, and BG appears  between all the lobes.
Note that, in the antiferromagnetic limit, the shrinking of the MI lobes 
is almost independent on the occupation and depends only on $U_2$.
In th ferromagnetic case, the MI lobes become more unstable increasing $n$ disappearing for 
for $n>(U_0+U_2+\Delta)/(2 \Delta)$.

So far we have analysed the effects of fluctuations along vertical and horizontal directions
in the phase plane. More generally, a diagonal disorder can be interpreted
in terms of fluctuations of a certain amplitude $\tilde{\Delta}$ along some direction, possibly
depending on the position  $(\mu^0,U^0_2)$ in the atomic phase diagram, in which is centred. 
Let us denote the direction as a straight line
\begin{equation}\label{eqn:straight}
 \frac{U_2}{U_0}=a\frac{\mu}{U_0}+b,
\end{equation}

In general, if a MI boundary lies along the direction of fluctuations defined by a specific
type of disorder, no BG would appear for sufficiently small disorder strength.

In the case of disorder  in $U_0$,  
the fluctuations occur along the direction defined by the parameters in (\ref{eqn:straight})
$a=U_2^0/\mu^0$ and $b=0$ with amplitude 
$\tilde{\Delta}=2 \Delta \sqrt{(\mu^0)^2+(U^0_2)^2} /(U_0-\Delta^2)$. 
The new boundaries are obtained  modifying $\mu_\pm$ by the replacements
$U_0\rightarrow U_0 \mp \Delta $.
This causes a progressive shrinking of all the lobes which disappear for high densities. 
It is worth noting that odd lobes disappear independently 
on $U_2$ for $n>\frac{U_0+\Delta}{2 \Delta}$. 
The resulting atomic phase diagram is shown in the bottom panel of 
Fig. \ref{fig:t0_disU2U0}.

\begin{figure}
\begin{center}
\includegraphics[width=0.55\linewidth,keepaspectratio,angle=-90]{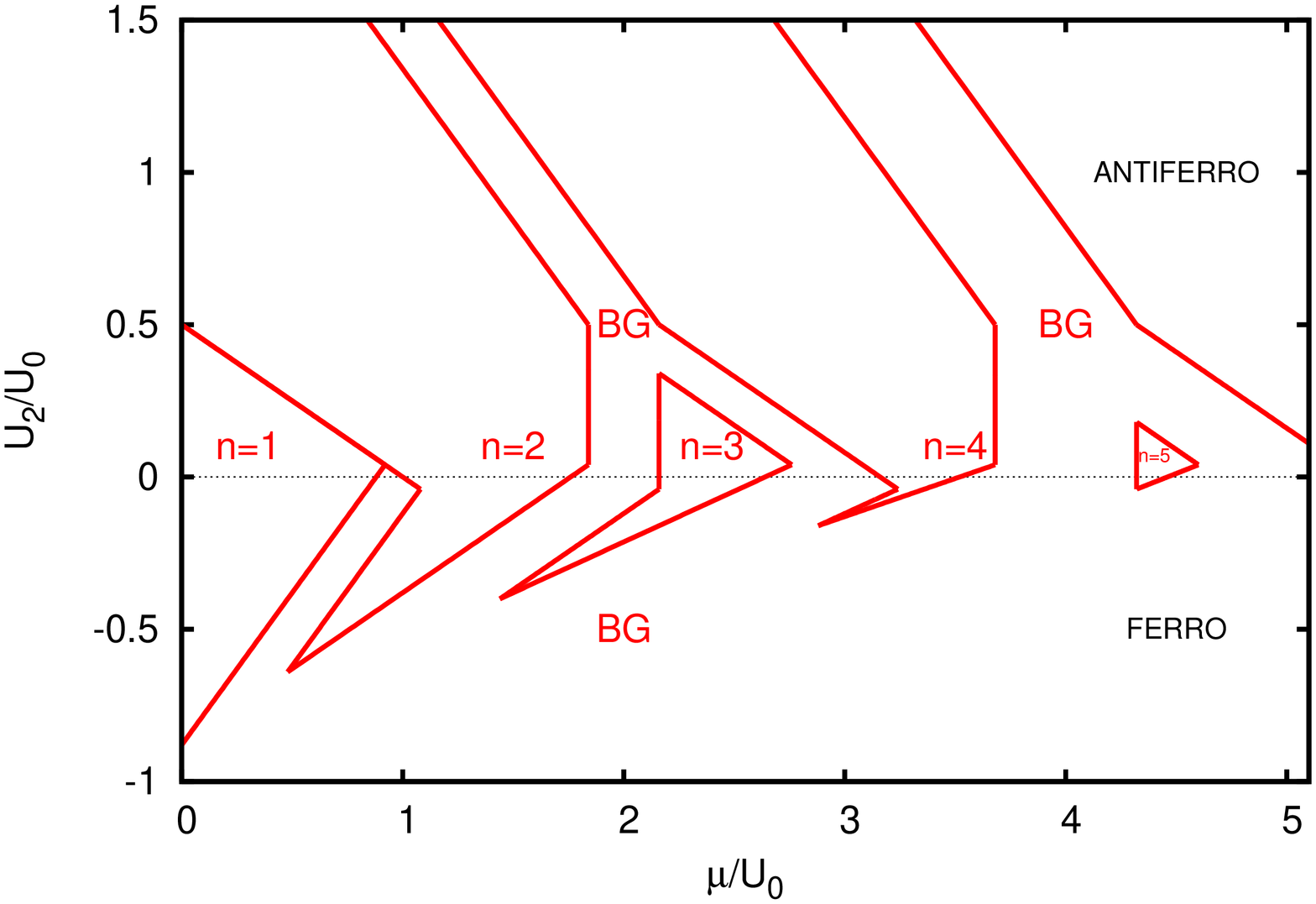}\\
\includegraphics[width=0.55\linewidth,keepaspectratio,angle=-90]{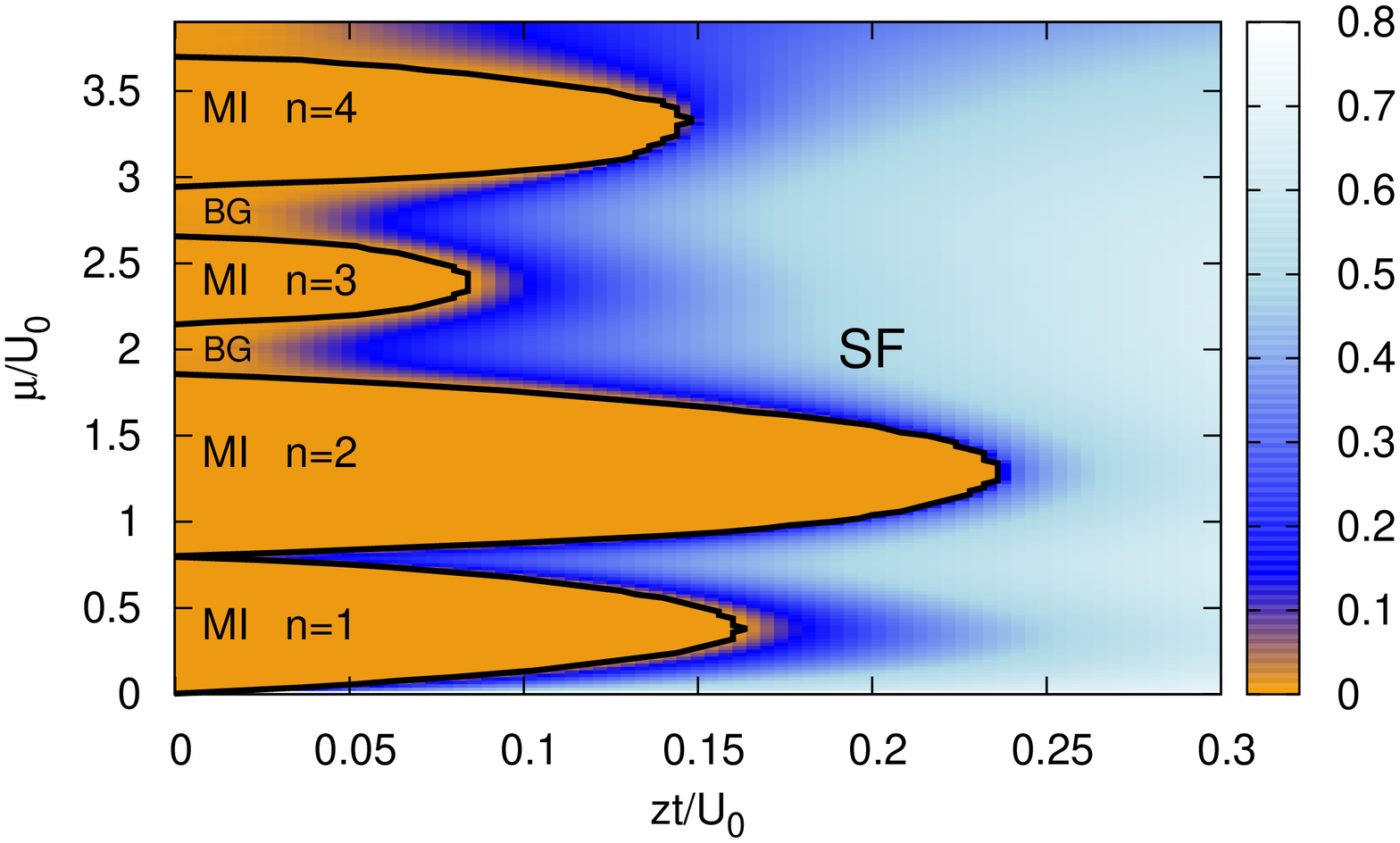}
\end{center}
\caption{(Color online) Disorder in $\alpha_2$ with $\Delta/U_0=0.04$. Top panel: atomic limit. 
Bottom panel: finite hopping phase diagram with $U_2/U_0=0.1$.}
\label{fig:disscat2}
\end{figure}

\begin{figure}
\begin{center}
\includegraphics[width=0.55\linewidth,keepaspectratio,angle=-90]{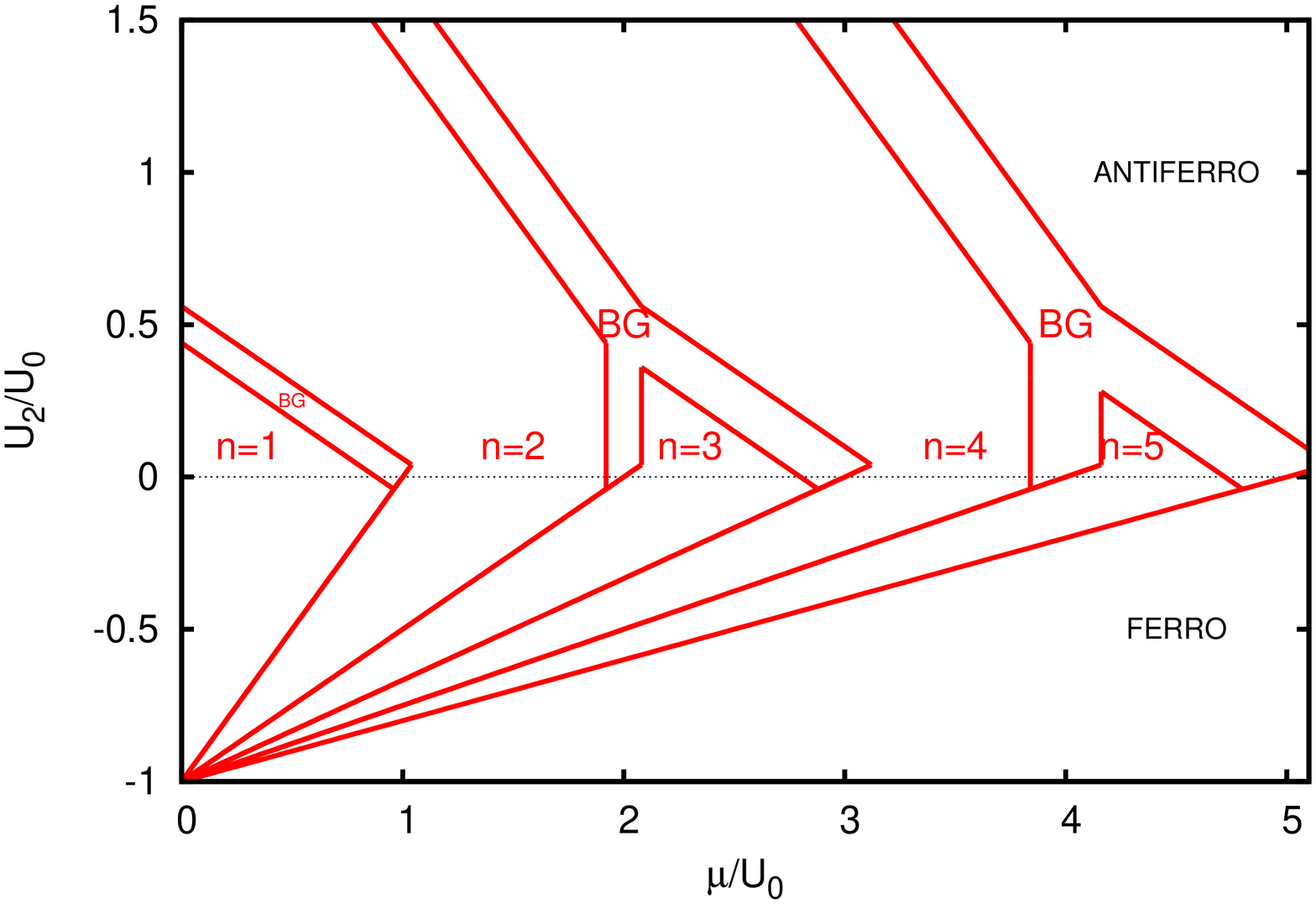}\\
\includegraphics[width=0.55\linewidth,keepaspectratio,angle=-90]{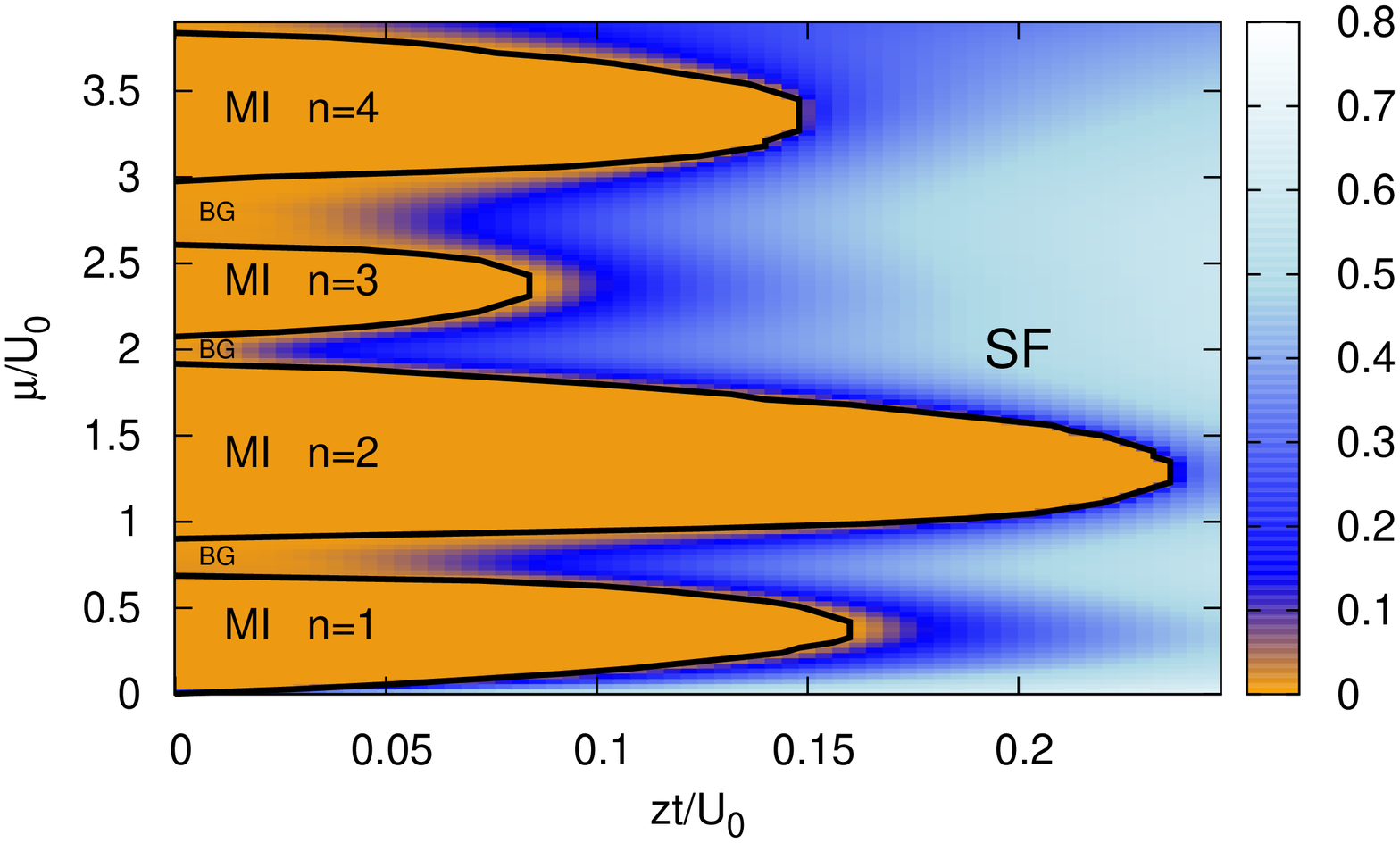}
\end{center}
\caption{(Color online) Disorder in $\alpha_0$ with $\Delta/U_0=0.04$. Top panel: atomic limit. 
Bottom panel: finite hopping phase diagram with $U_2/U_0=0.1$.}
\label{fig:disscat0}
\end{figure}
\section{Disorder in the scattering lengths}\label{sec:scat}

As pointed out in Sec. \ref{sec:model}, $U_0$ and $U_2$ depend 
 on the scattering lengths $a_0$ and $a_2$ via the relations 
\begin{eqnarray}
 U_0 &=& \alpha_0+2 \alpha_2, \nonumber \\
U_2 &=& \alpha_0- \alpha_2,
\end{eqnarray}
having introduced the renormalized parameters 
$\alpha_s =a_s 4 \pi \int d\vec{r} w^4(\vec{r}-\vec{r}_i)/(3m) $.
Both $\alpha_0$ and $\alpha_2$ can  fluctuate locally in the presence of optical Feshbach resonances,
so disorder can be introduced in these variables  and the new scattering lengths become  
site-dependent $\alpha^j_s=\alpha_s+\epsilon_j$.
This represents a more realistic scenario than considering disorder in $U_0$ or $U_2$ alone.

To find the phase diagram in the atomic limit, we have to modify again the boundaries in 
(\ref{eqn:bound1}-\ref{eqn:bound4}) as we did in the previous section in the cases of disorder
in $U_0$ and $U_2$, remembering that, this time, both of them fluctuate at the same time.
So, in the case of disorder in $\alpha_2$, one has to do the replacement
$U_0\rightarrow U_0 \mp 2 \Delta $ and $U_2\rightarrow U_2 \mp \Delta $  for $U_2/U_0>0$ and
$U_2\rightarrow U_2 \pm \Delta $  for $U_2/U_0<0$.
The fluctuations occur, with amplitude 
$\tilde{\Delta}=2 \Delta  \sqrt{(2\mu^0)^2+(U_0-2U^0_2)^2}/(U_0-4\Delta^2)$, along 
the direction (\ref{eqn:straight}) with $a=(U_2^0+U_0)/\mu_0$
and $b=0.5$ which is a family of straight lines passing through $\left(\mu/U_0=0,U_2/U_0=0.5\right)$. This means that the 
boundary between first and second lobe belongs to this family and no BG is expected between
these two MI regions, as can be observed in both the atomic-limit (top panel)
and complete phase (bottom panel) diagram in Fig. \ref{fig:disscat2}.

If the disorder is set in  $\alpha_0$, substitutions in Eqs. (\ref{eqn:bound1}-\ref{eqn:bound4})
are $U_0\rightarrow U_0 \mp \Delta $ and $U_2\rightarrow U_2 \pm \Delta $ for $U_2/U_0>0$ and
$U_s\rightarrow U_s \mp \Delta $ ($s=0,1$) for $U_2/U_0<0$. The direction of the fluctuations is given by
$a=(U_2^0+U_0)/\mu_0$ and $b=-1$ and amplitude 
$\tilde{\Delta}=2 \Delta \sqrt{ (\mu^0)^2+(U_0+U^0_2)^2}/(U_0-\Delta^2) $.
In this case, all the boundaries in the ferromagnetic limits lie on fluctuation directions so
no BG appears in the ferromagnetic regime as is shown in the top panel of Fig. \ref{fig:disscat0}.
This result marks a distinction between the scalar case and the spinor one  with ferromagnetic spin correlations
where disorder in the $s=0$ scattering channel only is not enough to produce BG. 
The finite-hopping phase diagram for the antiferromagnetic regime where MI lobes are always surrounded by BG is shown 
in the lower panel of  Fig.\ref{fig:disscat0}.

\section{Conclusions}\label{sec:conc}
We have analysed the spin-1 Bose Hubbard model with different types of diagonal disorder, 
focusing on the atomic limit to illustrate how Bose Glass  phase emerges in between Mott lobes.
In this limit, disorder mixes MI phases with different occupation numbers, producing regions of BG between MI intervals.
The study of the atomic limit gives useful information also about the finite hopping case, providing 
the structure of the phase diagram near the basis of the MI lobes. To illustrate the power of this approach, we have also shown
in some cases the complete phase diagram, calculated by  Gutzwiller approximation.
We first analysed disorder, in either the chemical potential or the local interaction $U_0$ and $U_2$, explaining
how to construct the phase diagram in the atomic limit. 
Then, we used these results to study the more realistic case of disorder in one of the two scattering lengths 
$a_0$ or $a_2$  corresponding to different
scattering channels.
While in the scalar case BG always appears between MI regions, as soon as a small disorder is introduced, the spinor character 
can stabilise the MI phase for some densities or spin interactions inhibiting the BG creation near its boundary.
That happens in the case of disorder in $U_2$ between lobes with $n=2m$ and $n=2m+1$, for disorder in 
$a_0$ in the ferromagnetic limit corresponding to $U_2/U_0<0$ as well as for disorder in  $a_2$ between the first and second lobe.
The creation of singlets in the even occupation MI enhances the  stability of  this phase, reducing 
the odd occupation MI lobes. As a consequence, the odd lobes are also less stable under the effect of disorder. 
In the extreme case in which odd lobes disappear also without disorder, the BG assumes a spin structure of 
singlet. 
As a future perspective, we would like to extend the approach  we have  illustrated in this paper to better enter 
into the  spin properties of BG phases as well as the MI in the regimes where ferromagnetic and antiferromagnetic orders are
mixed by disorder without destroying the MI phase.
In this limit some new interesting phases could appear where the glassy character is not embedded in the 
density but in the spin degrees of freedom.

\begin{acknowledgements}
We thank  M. Lewenstein  for useful discussions.  
Support from Polish Government (via research projects N202 079135 for 2008-2011 (M\L{}) and N202 124736 for 2009-2012 (JZ)), Spanish  Government (FIS2008:01236;02425, 
Consolider Ingenio 2010 (CDS2006-00019))
and Catalan Government (SGR2009:00347;00343) is acknowledged.
J. Z. acknowledges hospitality from ICFO and
partial support from the advanced ERC-grant QUAGATUA.
M.\L{}. acknowledges support from  
Jagiellonian University International Ph.D Studies in Physics of Complex 
Systems (Agreement no. MPD/2009/6) provided by  Foundation for Polish Science and cofinanced by the 
European Regional Development Funds.
S.P. is supported by the Spanish Ministry of Science
and Innovation through the program Juan de la Cierva. 
Computer simulations were performed at ACK Cyfronet AGH as a part of the POIG PL-Grid project (M.\L{}).
\end{acknowledgements}

\pagebreak

\end{document}